\documentclass[a4paper]{jpconf}
\usepackage{graphicx}
\usepackage{amssymb}
\usepackage{amsmath}
\usepackage{latexsym}
\newcommand{\Slash}[1]{{\ooalign{\hfil/\hfil\crcr$#1$}}}

\begin{document}
\title{A calculation for vector dark matter direct detection\footnote{
This talk is based on the work with Junji Hisano, Koji Ishiwata, and
Masato Yamanaka \cite{Hisano:2010yh}.
}}

\author{Natsumi Nagata$^{1, 2}$}

\address{$^1$ Department of Physics, Nagoya University, Nagoya 464-8602,
Japan}
\address{$^2$ Department of Physics, University of Tokyo, Tokyo 113-0033,
Japan}

\ead{natsumi@eken.phys.nagoya-u.ac.jp}

\begin{abstract}

 We evaluate the elastic scattering cross section of vector dark matter
 with nucleon based on the method of effective field theory. The dark
 matter is assumed to behave as a vector particle under the Lorentz
 transformation and to interact with colored particles including quarks in
 the Standard Model. After formulating general formulae for the
 scattering cross sections, we apply them to the case of the first
 Kaluza-Klein photon dark matter in the minimal universal extra
 dimension model. The resultant cross sections are found to be larger
 than those calculated in previous literature.

\end{abstract}

%%%%%%%%%%%%%%%%%%%%%%%%%%%%
\section{Introduction}
%%%%%%%%%%%%%%%%%%%%%%%%%%%%%

The existence of dark matter (DM) has been established by
cosmological observations \cite{Komatsu:2010fb}. One of the most
attractive candidates is what we call Weakly Interacting Massive
Particles (WIMPs), which are stable particles with masses of the
electroweak scale and weakly interact with ordinary matters. This
interactions enable us to search for WIMP DM by using the scattering
signal of DM with nuclei on the earth. Such kind of experiments are
called the direct detection experiments of WIMP DM.

For the past years, a lot of efforts have been dedicated to
the direct detection of WIMP DM, and their sensitivities have been
extremely improving. The XENON100 Collaboration, for example, 
gives a severe constraint on the spin-independent (SI) elastic scattering
cross section of WIMP DM with nucleon $\sigma^{\rm SI}_N$ ($\sigma^{\rm
SI}_N < 2.0\times 10^{-45}~{\rm cm}^2$ for WIMPs with a mass of
55~GeV$/c^2$)  \cite{Aprile:2012nq}. Moreover, ton-scale detectors for the
direct detection experiments are now planned and expected to have
significantly improved sensitivities.

In order to study the nature of DM based on these experiments, we need
to evaluate the WIMP-nucleon elastic scattering cross section precisely. 
In this work, we assume the WIMP DM to be a vector particle, and
evaluate its cross section scattering off a nucleon. Several candidates
for vector DM have been proposed in various models, and there
have been a lot of previous work computing the scattering cross
sections \cite{Cheng:2002ej, Servant:2002hb, Birkedal:2006fz}. However,
we found that in the calculations some of the leading contributions to
the scattering cross section are not evaluated correctly, or in some
cases completely neglected. Taking such situation into account, we study
the way of evaluating the cross section systematically by using the
method of effective field theory.

%%%%%%%%%%%%%%%%%%%%%%%%%%%%%%%%%%%%%%%%
\section{Direct detection of vector dark matter}
%%%%%%%%%%%%%%%%%%%%%%%%%%%%%%%%%%%%%%%%%%%%%%%%

In this section we discuss the way of evaluating the elastic scattering
cross section of vector DM with nucleon. First, we write down
the effective interactions of vector DM ($B_\mu$) with light quarks and
gluon \cite{Hisano:2010yh}: 
\begin{equation}
\mathcal{L}^{\mathrm{eff}}=\sum_{q=u,d,s}\mathcal{L}^{\mathrm{eff}}_q
+\mathcal{L}^{\mathrm{eff}}_G, 
\end{equation}
with
\begin{eqnarray}
\mathcal{L}^{\mathrm{eff}}_q &=&
  f_q m_q B^{\mu}B_{\mu}\bar{q}q+
\frac{d_q}{M}
  \epsilon_{\mu\nu\rho\sigma}B^{\mu}i\partial^{\nu}B^{\rho}
  \bar{q}\gamma^{\sigma}\gamma^{5}q+\frac{g_q}{M^2}
  B^{\rho}i\partial^{\mu}i\partial^{\nu}B_{\rho}\mathcal{O}^q_{\mu\nu},
 \label{eff_lagq}
\\
\mathcal{L}^{\mathrm{eff}}_G&=&f_G
 B^{\rho}B_{\rho}G^{a\mu\nu}G^a_{\mu\nu},
\label{eff_lagG}
\end{eqnarray} 
where $m_q$ are the masses of light quarks, $M$ is the DM mass, and
$\epsilon^{\mu\nu\rho\sigma}$ is the totally antisymmetric tensor
defined as $\epsilon^{0123}=+1$. The covariant derivative is
defined as $D_\mu\equiv\partial_\mu+i g_sA^a_\mu T_a$, with $g_s$, $T_a$
and $A^a_\mu$ being the strong coupling constant, the SU(3)$_C$
generators, and the gluon fields, respectively. The gluon field strength
tensor is denoted by $G^a_{\mu\nu}$, and
$\mathcal{O}^q_{\mu\nu}\equiv\frac12 \bar{q} i \left(D_{\mu}\gamma_{\nu}
+ D_{\nu}\gamma_{\mu} -\frac{1}{2}g_{\mu\nu}\Slash{D} \right) q $ are the
twist-2 operators of light quarks. When we write down the effective
Lagrangian, we consider the fact that the scattering process is
non-relativistic. The coefficients of the operators are to be determined
by integrating out the heavy particles in high energy theory. The second
term in Eq.~(\ref{eff_lagq}) gives rise to the spin-dependent (SD)
interaction, while the other terms yield the spin-independent (SI)
interactions. We focus on the SI interactions hereafter,
because the experimental constraint is much severe for the
SI interactions, rather than for the SD interactions.

In order to obtain the effective coupling of the vector DM with
nucleon induced by the effective Lagrangian, we need to evaluate the
nucleon matrix elements of the quark and gluon operators in
Eqs.(\ref{eff_lagq}) and (\ref{eff_lagG}). First, the nucleon matrix
elements of the scalar-type quark operators are parametrized as
\begin{equation}
 f_{Tq}\equiv \langle N \vert m_q \bar{q} q \vert N\rangle/m_N~,
\end{equation}
with $\vert N\rangle$ and $m_N$ the one-particle state and the mass of
nucleon, respectively. The parameters are called the mass fractions and
their values are obtained from the lattice simulations
\cite{Young:2009zb, :2012sa}. Second, for the quark twist-2 operators,
we can use the parton distribution functions (PDFs):
\begin{eqnarray}
\langle N(p)\vert 
{\cal O}_{\mu\nu}^q
\vert N(p) \rangle 
&=&\frac{1}{m_N}
(p_{\mu}p_{\nu}-\frac{1}{4}m^2_N g_{\mu\nu})\
(q(2)+\bar{q}(2)) \ ,
\end{eqnarray}
where $q(2)$ and $\bar{q}(2)$ are the second moments of PDFs of quark
$q(x)$ and anti-quark $\bar{q}(x)$, respectively, which are defined as
$q(2)+ \bar{q}(2)  =\int^{1}_{0} dx ~x~ [q(x)+\bar{q}(x)]$. These values
are obtained from Ref.~\cite{Pumplin:2002vw}. Finally, the matrix element
of gluon field strength tensor can be evaluated by using the trace
anomaly of the energy-momentum tensor in QCD \cite{Shifman:1978zn}. The
resultant expression is given as
\begin{equation}
 \langle N\vert G^a_{\mu\nu}G^{a\mu\nu}\vert N\rangle
=-\frac{8\pi}{9\alpha_s} m_N f_{TG}
\end{equation}
with $f_{TG}\equiv 1-\sum_{q=u,d,s}f_{Tq}$. Note that the right hand
side of the expression is divided by the strong coupling constant,
$\alpha_s$.  For this reason, although the gluon contribution is induced
by higher loop diagrams, it can be comparable to the quark contributions
\cite{Hisano:2010ct}. Briefly speaking, the enhancement comes from the
large gluon contribution to the mass of nucleon. As a result, the SI
effective coupling of vector DM with nucleon, $f_N$, is given as
\begin{eqnarray}
f_N/m_N&=&\sum_{q=u,d,s}
f_q f_{Tq}
+\sum_{q=u,d,s,c,b}
\frac{3}{4} \left(q(2)+\bar{q}(2)\right)g_q
-\frac{8\pi}{9\alpha_s}f_{TG} f_G ~. 
\label{f}
\end{eqnarray}

Using the effective coupling, we eventually obtain the SI scattering
cross section of DM with nucleon: 
\begin{equation}
 \sigma^{\rm (SI)}_{N}=
\frac{1}{\pi}\biggl(\frac{m_N}{M+m_N}\biggr)^2~\vert f_N\vert ^2~.
\end{equation}

Now, all we have to do reduces to evaluate the coefficients of the
effective operators by integrating out the heavy fields in the
high-energy theories. For example, we take the case where the
interaction Lagrangian of the vector DM has a generic form as
\begin{eqnarray}
  \mathcal{L}=
  \bar{\psi}_2 ~(a_{\psi_2\psi_1} \gamma^{\mu}+b_{\psi_2\psi_1}
  \gamma^{\mu}\gamma_5)\psi_1  B_{\mu}   + \mathrm{h.c.}~,
\label{eq:simpleL}
\end{eqnarray}
where $\psi_1$ and $\psi_2$ are colored fermions with masses $m_1$ and
$m_2$ ($m_1<m_2$), respectively.
%
%%%%%%%%%%%%%%%%%%FIGURE%%%%%%%%%%%%%%%%%%%
\begin{figure}[t]
 \begin{center}
  \includegraphics[height=4cm]{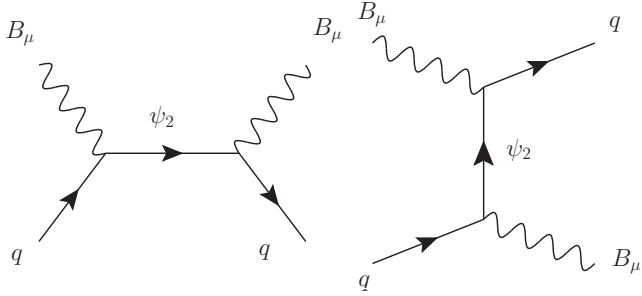}
  \caption{Tree-level diagrams of exchanging colored fermion $\psi_2$
     to generate interaction of vector dark matter with
    light quarks.}
\label{fig:tree_general}
 \end{center}
\end{figure}
%%%%%%%%%%%%%%%%%%%%%%%%%%%%%%%%%%%
%
In this case, the vector DM is scattered by light quarks at tree-level. The
relevant interaction Lagrangian is given by taking $\psi_1=q$ in
Eq.~(\ref{eq:simpleL}), and the corresponding diagrams are shown in
Fig.~\ref{fig:tree_general}.  After integrating out the heavy particle
$\psi_2$, we obtain
\begin{eqnarray}
f_q&=&\frac{a^2_{\psi_2 q}-b^2_{\psi_2 q}}{m_q}\frac{m_{2}}{m^2_{2}-M^2}
-(a^2_{\psi_2 q}+b^2_{\psi_2 q})
\frac{m^2_{2}}{2(m^2_{2}-M^2)^2},   
\label{tree_general_1}  \\
g_q&=&-\frac{2M^2(a^2_{\psi_2 q}+b^2_{\psi_2 q})}{(m^2_{2}-M^2)^2}.
\label{tree_general_3}
\end{eqnarray}
One can easily find that the effective
couplings obtained here are enhanced when the vector DM and
the heavy colored fermion are degenerate in mass \cite{Hisano:2011um}.

%%%%%%%%%%%%%%%%%%FIGURE%%%%%%%%%%%%%%%%%%%
\begin{figure}[t]
 \begin{center}
  \includegraphics[height=4cm]{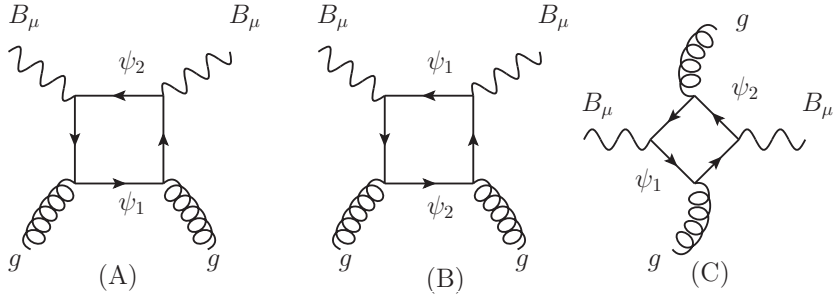}
  \caption{One-loop contributions to scalar-type effective
    coupling with gluon. }
\label{fig:loop_general}
 \end{center}
\end{figure}
%%%%%%%%%%%%%%%%%%%%%%%%%%%%%%%%%%%

The effective coupling of the vector DM with gluon is induced by 1-loop
diagrams illustrated in Fig.~\ref{fig:loop_general}.  In those
processes, all the particles $\psi_1$ and $\psi_2$
which couple with $B_{\mu}$ run in the loop. The resultant expressions
are somewhat complicated, and thus we just quote
Ref.~\cite{Hisano:2010yh} for their complete formulae as well as
their derivation.

%%%%%%%%%%%%%%%%%%%%%%%%%%%%%%%%%%%%%%%
\section{Application and Results}
%%%%%%%%%%%%%%%%%%%%%%%%%%%%%%%%%%%%%%%

Next, we deal with a particular model for vector DM as an
application. We carry out the calculation for the first Kaluza-Klein
(KK) photon DM \cite{Cheng:2002ej, Servant:2002hb} in the minimal
universal extra dimension (MUED) model \cite{Appelquist:2000nn,
Cheng:2002iz}. In this model, an extra dimension is compactified on an
$S^1/\mathbb{Z}_2$ orbifold with the compactification radius $R$, and all
of the Standard Model (SM) particles propagate in the dimension. The
lightest KK-odd particle (LKP) is prevented from decaying to the SM
particles, so it becomes DM. This model has just three undetermined
parameters: the radius of the extra dimension $R$, the mass of Higgs
boson $m_h$, and the cutoff scale $\Lambda$.

In general, extra dimensional models give rise to the degenerate mass
spectrum at tree-level, which is broken by radiative corrections.
By evaluating the radiative corrections \cite{Cheng:2002iz}, one
finds that the first KK photon $B^{(1)}$ is the lightest KK-odd particle,
thus, becomes the DM in the Universe. Moreover, since the mass difference is
induced by radiative corrections, the mass degeneracy is tight for a
small cut-off scale. In such a case, although it is difficult to probe
the MUED model at the LHC because of the soft QCD jets, the direct
detection rate of dark matter is expected to be enhanced
\cite{Hisano:2011um, Arrenberg:2008wy}, as we shall see soon
later\footnote{
Indirect DM searches also might be a powerful alternative in such a case
\cite{Asano:2011ik, Garny:2012eb}. 
}.

%%%%%%%%%%%%%%%%%%FIGURE%%%%%%%%%%%%%%%%%%%
\begin{figure}
 \begin{center}
   \includegraphics[height=3.5cm]{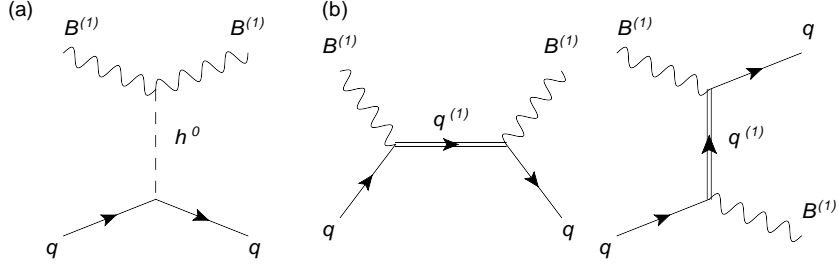}
   \caption{Tree-level diagrams for the elastic scattering of
  the KK photon DM
     $B^{(1)}$ with light quarks: (a) Higgs boson exchange
     contribution, and (b) KK quark exchange contributions.}
\label{fig:tree}
 \end{center}
\end{figure}
%%%%%%%%%%%%%%%%%%%%%%%%%%%%%%%%%%%
Now we evaluate the SI scattering cross section of the KK
photon DM with nucleon. The effective interaction of the KK
photon DM $B^{(1)}$ with light quarks is induced by the tree-level
diagrams shown in Fig.~\ref{fig:tree}. Here, $h^0$ and $q^{(1)}$ are the
Higgs boson and the first KK quark, respectively. 
%%%%%%%%%%%%%%%%%%FIGURE%%%%%%%%%%%%%%%%%%%
\begin{figure}
 \begin{center}
   \includegraphics[height=4cm]{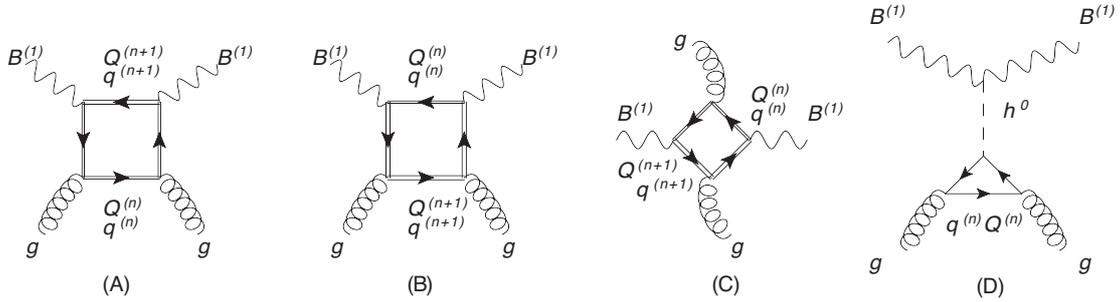}
   \caption{One-loop diagrams for the effective interaction of
     $B^{(1)}$ with gluon.}
\label{fig:qloop}
 \end{center}
\end{figure}
%%%%%%%%%%%%%%%%%%%%%%%%%%%%%%%%%%%
Also, there are one-loop diagrams we should evaluate for the gluon
contribution. They are illustrated in Fig.~\ref{fig:qloop}. In these
diagrams, all of the KK quarks run in the loop. Taking all the
contributions into account, we obtain the effective interactions. Their
expressions are given in Ref.~\cite{Hisano:2010yh}.

%%%%%%%%%%%%%%%%%%%%%%%%%%%%%%%%%%%%%%%%%
\begin{figure}[t]
 \begin{center}
   \includegraphics[height=7cm]{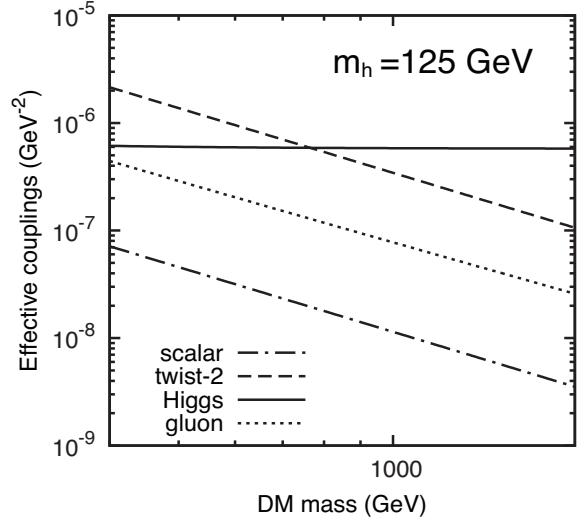}
   \caption{Each contribution in effective coupling $f_N/m_N$ given in
     Eq.~(\ref{f}). Here we set $m_h=125~{\rm GeV}$ and $(m_{\rm
       1st}-M)/M=0.1$.}
\label{fig:fN_10}
 \end{center}
\end{figure}
%%%%%%%%%%%%%%%%%%%%%%%%%%%%%%%%%%%%%%%%%%%%%%%%

%%%%%%%%%%%%%%%%%%%%%%%%%%%%%%%%%%%%%%%%
\begin{figure}[t]
  \begin{center}
    \includegraphics[height=7.5cm]{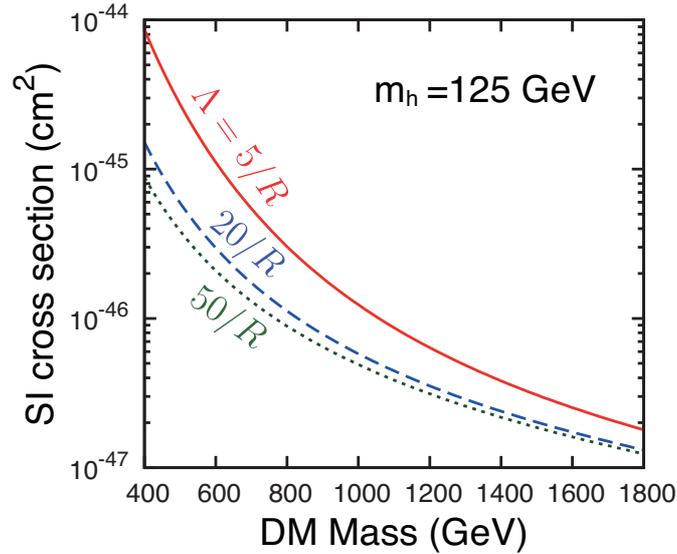}
  \end{center}
  \caption{Spin-independent cross section with proton for $m_h = 125$
    GeV. Each line corresponds to $\Lambda = 5/R$, $20/R$, and $50/R$,
    respectively. }
  \label{mhfixed}
\end{figure}
%%%%%%%%%%%%%%%%%%%%%%%%%%%%%%%%%%%%%%%%%%%%%%%%%%%%%%%%%%%
In Fig.~\ref{fig:fN_10}, we plot each contribution to the SI effective
coupling of DM with a proton. The solid line shows the contribution of
the Higgs boson exchanging diagrams, the dashed line indicates the
twist-2 type contribution, the dash-dotted line corresponds to the
scalar-type contribution (except for the Higgs-exchanging contribution),
and the dotted line represents the gluon contribution (except for the
Higgs-exchanging contribution). Here we set the Higgs boson
mass\footnote{Recent searches for the Standard Model Higgs boson at the
LHC indicate its mass to be around 125~GeV \cite{:2012gk, :2012gu}.} equal
to 125~GeV, and the
mass difference between the DM and the first KK quark to be 10 \% by hand. 
We find that all of the contributions have the same sign thus they are
constructive. The twist-2 contribution is dominant when $M\lesssim
800~{\rm GeV}$, while that of Higgs boson exchanging process becomes
dominant above it. Moreover, although the tree-level quark contributions
are dominant, it is found that the gluon contribution is not negligible
at all.

By using the effective couplings obtained above, we evaluate the
SI scattering cross sections. In Fig.~\ref{mhfixed}, we plot the
SI cross section of KK photon DM with a proton as a function of 
the DM mass. Here again, the Higgs boson mass is set to be 125
GeV. We take the cut-off scale $\Lambda=5/R,~20/R, ~50/R$ from top to
bottom. 
We find that the scattering cross sections reduce as the cut-off
scale is taken to be large, as expected.
As a result, we obtain the SI scattering cross section which is
larger than those obtained by the previous calculations. Note that the
DM mass with which the thermal relic abundance is preferred by the WMAP
observation \cite{Komatsu:2010fb} is around 1300~GeV
\cite{Belanger:2010yx}. It corresponds to $\sigma^{\rm
SI}_N=$(2.5-5)$\times 10^{-47}$~cm$^{2}$, so the direct detection
experiments with ton-scale detectors might be able to probe the DM in
the future.

%%%%%%%%%%%%%%%%%%%%%%%%%%%%%%%%%%%%%%%%
\section{Conclusion}
%%%%%%%%%%%%%%%%%%%%%%%%%%%%%%%%%%%%%%%%

We calculate the spin independent elastic scattering cross sections of
vector dark matter with nucleon based on the effective field theory. It is
found that the interaction of dark matter with gluon as well as quarks yields
sizable contribution to the scattering cross section, though the gluon
contribution is induced at loop level. 
The scattering cross section of the first Kaluza-Klein 
photon dark matter in the MUED model turns out to be larger than those
obtained by the previous calculations.

%%%%%%%%%%%%%%%%%%%%%%%%%%%%%%%%%%%%
\ack
%%%%%%%%%%%%%%%%%%%%%%%%%%%%%%%%%%%%

The author would like to thank Junji Hisano, Koji Ishiwata, and Masato
Yamanaka for collaboration.  
This work is supported by Research Fellowships of the Japan
Society for the Promotion of Science for Young Scientists.

%%%%%%%%%%%%%%%%%%%%%%%%%%%%%%%%%%%%%%%


\section*{References}
\begin{thebibliography}{99}
%%%%%%%%%%%%%%%%%%%%%%%%%%%%%%%%%%%%%%%%%


%\cite{Hisano:2010yh}
\bibitem{Hisano:2010yh} 
  J.~Hisano, K.~Ishiwata, N.~Nagata and M.~Yamanaka,
  %``Direct Detection of Vector Dark Matter,''
  Prog.\ Theor.\ Phys.\  {\bf 126}, 435 (2011)
%  [arXiv:1012.5455 [hep-ph]]
.
  %%CITATION = ARXIV:1012.5455;%%


%\cite{Komatsu:2010fb}
\bibitem{Komatsu:2010fb} 
  E.~Komatsu {\it et al.}  [WMAP Collaboration],
  %``Seven-Year Wilkinson Microwave Anisotropy Probe (WMAP) Observations: Cosmological Interpretation,''
  Astrophys.\ J.\ Suppl.\  {\bf 192}, 18 (2011)
%  [arXiv:1001.4538 [astro-ph.CO]]
.
  %%CITATION = ARXIV:1001.4538;%%

%\cite{Aprile:2012nq}
\bibitem{Aprile:2012nq} 
  E.~Aprile {\it et al.}  [XENON100 Collaboration],
  %``Dark Matter Results from 225 Live Days of XENON100 Data,''
  arXiv:1207.5988 [astro-ph.CO].
  %%CITATION = ARXIV:1207.5988;%%

\bibitem{Cheng:2002ej}
  H.~C.~P.~Cheng, J.~L.~Feng and K.~T.~Matchev,
  %``Kaluza-Klein dark matter,''
  Phys.\ Rev.\ Lett.\  {\bf 89}, 211301  (2002).
  %%CITATION = PRLTA,89,211301;%%

\bibitem{Servant:2002hb}
  G.~Servant and T.~M.~P.~Tait,
  %``Elastic scattering and direct detection of Kaluza-Klein dark matter,''
  New J.\ Phys.\  {\bf 4}, 99 (2002).
   %%CITATION = NJOPF,4,99;%%

\bibitem{Birkedal:2006fz}
  A.~Birkedal, A.~Noble, M.~Perelstein and A.~Spray,
  %``Little Higgs dark matter,''
  Phys.\ Rev.\  D {\bf 74}, 035002 (2006).

%\cite{Young:2009zb}
\bibitem{Young:2009zb} 
  R.~D.~Young and A.~W.~Thomas,
  %``Octet baryon masses and sigma terms from an SU(3) chiral extrapolation,''
  Phys.\ Rev.\ D {\bf 81}, 014503 (2010)
%  [arXiv:0901.3310 [hep-lat]]
.
  %%CITATION = ARXIV:0901.3310;%%

%\cite{:2012sa}
\bibitem{:2012sa} 
 H. Ohki , {\it et al.}  [JLQCD Collaboration],
  %``Nucleon strange quark content from N_f=2+1 lattice QCD with exact chiral symmetry,''
  arXiv:1208.4185 [hep-lat].
  %%CITATION = ARXIV:1208.4185;%%

\bibitem{Pumplin:2002vw}
  J.~Pumplin, D.~R.~Stump, J.~Huston, H.~L.~Lai, P.~M.~Nadolsky and W.~K.~Tung,
  %``New generation of parton distributions with uncertainties from global QCD
  %analysis,''
  JHEP {\bf 0207}, 012 (2002).

%\cite{Shifman:1978zn}
\bibitem{Shifman:1978zn} 
  M.~A.~Shifman, A.~I.~Vainshtein and V.~I.~Zakharov,
  %``Remarks on Higgs Boson Interactions with Nucleons,''
  Phys.\ Lett.\ B {\bf 78}, 443 (1978).
  %%CITATION = PHLTA,B78,443;%%

%\cite{Hisano:2010ct}
\bibitem{Hisano:2010ct} 
  J.~Hisano, K.~Ishiwata and N.~Nagata,
  %``Gluon contribution to the dark matter direct detection,''
  Phys.\ Rev.\ D {\bf 82}, 115007 (2010)
%  [arXiv:1007.2601 [hep-ph]]
.
  %%CITATION = ARXIV:1007.2601;%%

%\cite{Hisano:2011um}
\bibitem{Hisano:2011um} 
  J.~Hisano, K.~Ishiwata and N.~Nagata,
  %``Direct Detection of Dark Matter Degenerate with Colored Particles
	%in Mass,'' 
  Phys.\ Lett.\ B {\bf 706}, 208 (2011)
%  [arXiv:1110.3719 [hep-ph]]
.
  %%CITATION = ARXIV:1110.3719;%%

%\cite{Appelquist:2000nn}
\bibitem{Appelquist:2000nn} 
  T.~Appelquist, H.~-C.~Cheng and B.~A.~Dobrescu,
  %``Bounds on universal extra dimensions,''
  Phys.\ Rev.\ D {\bf 64}, 035002 (2001)
%  [hep-ph/0012100]
.
  %%CITATION = HEP-PH/0012100;%%

%\cite{Cheng:2002iz}
\bibitem{Cheng:2002iz} 
  H.~-C.~Cheng, K.~T.~Matchev and M.~Schmaltz,
  %``Radiative corrections to Kaluza-Klein masses,''
  Phys.\ Rev.\ D {\bf 66}, 036005 (2002)
%  [hep-ph/0204342]
.
  %%CITATION = HEP-PH/0204342;%%

%\cite{Arrenberg:2008wy}
\bibitem{Arrenberg:2008wy} 
  S.~Arrenberg, L.~Baudis, K.~Kong, K.~T.~Matchev and J.~Yoo,
  %``Kaluza-Klein Dark Matter: Direct Detection vis-a-vis LHC,''
  Phys.\ Rev.\ D {\bf 78}, 056002 (2008)
%  [arXiv:0805.4210 [hep-ph]]
.
  %%CITATION = ARXIV:0805.4210;%%

%\cite{Asano:2011ik}
\bibitem{Asano:2011ik} 
  M.~Asano, T.~Bringmann and C.~Weniger,
  %``Indirect dark matter searches as a probe of degenerate particle spectra,''
  Phys.\ Lett.\ B {\bf 709}, 128 (2012). 
%  [arXiv:1112.5158 [hep-ph]].
%  %%CITATION = ARXIV:1112.5158;%%
%
%\cite{Garny:2012eb}
\bibitem{Garny:2012eb} 
M.~Garny, A.~Ibarra, M.~Pato and S.~Vogl,
  %``Closing in on mass-degenerate dark matter scenarios with antiprotons and direct detection,''
  arXiv:1207.1431 [hep-ph].
  %%CITATION = ARXIV:1207.1431;%%

%\cite{:2012gk}
\bibitem{:2012gk} 
  G.~Aad {\it et al.}  [ATLAS Collaboration],
  %``Observation of a new particle in the search for the Standard Model Higgs boson with the ATLAS detector at the LHC,''
  Phys.\ Lett.\ B {\bf 716}, 1 (2012)
%  [arXiv:1207.7214 [hep-ex]]
.
  %%CITATION = ARXIV:1207.7214;%%

%\cite{:2012gu}
\bibitem{:2012gu} 
  S.~Chatrchyan {\it et al.}  [CMS Collaboration],
  %``Observation of a new boson at a mass of 125 GeV with the CMS experiment at the LHC,''
  Phys.\ Lett.\ B {\bf 716}, 30 (2012)
%  [arXiv:1207.7235 [hep-ex]]
.
  %%CITATION = ARXIV:1207.7235;%%

%\cite{Belanger:2010yx}
\bibitem{Belanger:2010yx} 
  G.~Belanger, M.~Kakizaki and A.~Pukhov,
  %``Dark matter in UED: The Role of the second KK level,''
  JCAP {\bf 1102}, 009 (2011)
%  [arXiv:1012.2577 [hep-ph]]
.
  %%CITATION = ARXIV:1012.2577;%%

\end{thebibliography}
\end{document}